\documentclass{PoS}
\usepackage{epsf,color}
\usepackage{latexsym}
\usepackage{amsmath}

\usepackage[normalem]{ulem}

\title{Structure and EM form factors of purely relativistic systems}

\ShortTitle{Structure of relativistic systems}

\author{\speaker{V.A.~Karmanov}\\
Lebedev Physical Institute, Leninsky prospect 53, 11991 Moscow, Russia\\
        E-mail: \email{karmanovva@lebedev.ru}}

\author{J.~Carbonell\\
IPNO, Univ. Paris-Sud, CNRS/IN2P3, Universit\'e Paris-Saclay,
  91405 Orsay, France\\
          E-mail: \email{carbonell@ipno.in2p3.fr}}
          
\author{H.~Sazdjian\\
IPNO, Univ. Paris-Sud, CNRS/IN2P3, Universit\'e Paris-Saclay,
  91405 Orsay, France\\
          E-mail: \email{sazdjian@ipno.in2p3.fr}}          

\abstract{The Bethe-Salpeter equation for two massive scalar particles
interacting by scalar massless exchange has solutions of two types,
which differ from each other by their   behavior in the non-relativistic
limit: the normal solutions which turn into the Coulomb ones 
and the "abnormal" solutions. The latter ones have no non-relativistic
counterparts and disappear in the non-relativistic limit. We studied
the composition of all these states. It turns out that the normal states,
even for large binding energy, are dominated by two massive particles.
Whereas, the contribution of the two-body sector into the abnormal
states, even for small binding energy, is of the order of 1\% only;
they are dominated by an indefinite number of the massless particles.
The elastic electromagnetic form factors for both normal and abnormal
states, as well as the transition ones between them, are calculated.}

\FullConference{Light Cone 2019 - QCD on the light cone:
from hadrons to heavy ions - LC2019\\
16-20 September 2019\\
Ecole Polytechnique, Palaiseau, France}

\begin{document}

\section{Introduction}\label{intro}
In the Wick-Cutkosky model \cite{wick,cutk} (two scalar constituent
particles interacting by massless scalar exchange) and for enough large
coupling constant ($\alpha>\frac{\pi}{4}$), there exist two different
types of solutions of the Bethe-Salpeter (BS) equation \cite{bs}. 
In the non-relativistic limit (understood as the speed of light $c$,
taken as a parameter, tending to infinity: $c\to\infty$),
some solutions  $\Phi(k,p)$
turn  into the well-known solutions in the Coulomb potential;
other solutions, on the contrary,  disappear. 
The latter ones
have purely relativistic origin and are called "abnormal". They can
be odd or even relative to $k_0\to -k_0$. However, as shown in
\cite{cia}, the odd solutions do not contribute in the S-matrix and
therefore, they are hardly physical ("observable"). 
The search for the abnormal symmetric states generated by massive exchange is
in progress. 

The following questions immediately arise: 
Are the abnormal states a consequence of a mathematical flaw of the
BS equation? If not, how are they constructed, i.e. which constituents do
they contain? Why do they not have non-relativistic counterparts? 

In this contribution, still in the Wick-Cutkosky model, i.e., for the
massless exchange, we will study  the physical nature of the abnormal
states and answer these questions. 

%%%%%%%%%%%%%%%%%%%%%%%%%%%%%%%%%%%%%
\section{Two-body contribution and EM form factors}

The state vector $|p\rangle$ is schematically represented via the Fock
decomposition as follows:
\begin{equation}\label{eq1}
|p\rangle=\psi_2 |2\rangle+\psi_3 |3\rangle+\psi_4 |4\rangle+\ldots,
\end{equation}
where in the Wick-Cutkosky model the decomposition starts with the
two-body state $|2\rangle$ containing two constituent particles only
(each with mass $m$). 
Each higher state still contains two constituent particles, but with
extra exchanged particles.
The state vector $|p\rangle$ is normalized as
\begin{equation}\label{eq2}
\langle p|p\rangle=N_2+N_3+\ldots=1,
\end{equation}
where $N_2$, $N_3$ etc. are contributions of the two-, three-, etc.
Fock sectors.
Using the BS solution, we will be able to find  $\psi_2$ and,
correspondingly, $N_2$. To this aim, we will use the  relation between
the BS amplitude and $\psi_2$. It has the form (see, e.g.,
\cite{cdkm}, Sec. 3.3):
\begin{equation}\label{eq3}
\psi_2(\vec{k}_{\perp},x)=\frac{
(\frac{1}{2}\omega\cdot p+\omega\cdot k)
(\frac{1}{2}\omega\cdot p-\omega\cdot k)}{\pi(\omega\cdot p)}
\int_{-\infty}^{\infty}\Phi(k+\beta\omega, p)d\beta\ .
\end{equation}
We use here the explicitly covariant version of light-front dynamics
\cite{cdkm}, where the state vector is defined on the hyperplane 
$\omega\cdot x=0$, taking for $\omega=(\omega_0,\vec{\omega})$ the
four-vector with $\omega^2=0$, with an arbitrary orientation of 
$\vec{\omega}$. 
The standard version corresponds to the particular case
$\omega=(1,0,0,-1)$.
The BS amplitude $\Phi$ is normalized by the condition that the
electromagnetic (em) form factor, expressed through it, is
equal to 1 at $q=0$. This normalization is just equivalent to the condition
(\ref{eq2}).

This method was used in \cite{dshvk} to find the dependence of the
two-body contribution $N_2$ to the normal ground state on the coupling
constant $\alpha$. It was found that for small $\alpha$ (small binding
energy $B\ll m$, total mass $M\approx 2m$) $N_2$ is very close to 1.
For large $\alpha$ ($\alpha\to 2\pi$, the critical value, providing
the solution with $M^2\to 0$  and the binding energy $B\to 2m$) $N_2$
tends to $9/14\approx 64\%$. Our consideration here is completely
analogous to what was done in \cite{dshvk}. The only difference is
that instead of the normal BS solution $\Phi(k,p)$, we take
the abnormal one.

The BS solution is represented via the weight function $g(z)$ in the
form of Nakanishi integral representation \cite{N_63}
\begin{equation}\label{naka}
\Phi(k;p)=\frac{i}{\sqrt{N_{tot}}}\int_{-1}^1
\frac{g(z)dz}{\left(k^2+p\cdot k\; z+\frac{1}{4}M^2-m^2
+i\epsilon\right)^3}\ .
\end{equation}
The normalization factor $N_{tot}$ will be found below. 

The function $g(z)$ satisfies a differential equation  \cite{wick,cutk}
and is labelled by two quantum numbers $n,k$. The states with $k=0$ are
the normal ones.  In the non-relativistic limit the binding energies
reproduce the Coulomb spectrum $B_n=2m-M=\frac{\alpha^2m}{4n^2}$.
The states with $k\neq 0$ are the abnormal ones. The value of $k$
corresponds to the number of nodes of the function $g(z)$. For even
$k$ the function $g(z)$ is symmetric relative to $z\to -z$, for odd $k$
it is antisymmetric. We will consider the solutions with $n=1$ and
$k=0,2,4$. Then the equation for $g(z)$ obtains the form:
\begin{equation}\label{g}
g''(z)+\frac{\alpha}{\pi}\frac{1}{(1-z^2)(1-\eta^2+\eta^2z^2)}g(z)=0,
\end{equation}
where $\eta=\frac{M}{2m}=1-\frac{B}{2m}$ and the boundary conditions
are $g(\pm 1)=0$. We solved this equation numerically, generating
the spectrum labelled by $k$.

Substituting $\Phi(k;p)$ from (\ref{naka}) into (\ref{eq3}), we find
$\psi_2(\vec{k}_{\perp},x)$ in terms of $g(z)$:
\begin{equation}\label{eq4}
\psi(\vec{k}_{\perp},x)=\frac{1}{\sqrt{N_{tot}}}\frac{x(1-x)g(1-2x)}
{\Bigl[\vec{k}_{\perp}^2+m^2-x(1-x)M^2\Bigr]^2}\ .
\end{equation}
In this way, we find the two-body contribution into the full norm
(equaled to 1):
\begin{equation}\label{norm2}
N_2=\langle 2|2\rangle=\frac{1}{(2\pi)^3}\int\psi^2_n(\vec{k}_{\perp},x)
\frac{d^2k_{\perp}dx}{2x(1-x)}=\frac{1}{48\pi^2N_{tot}}
\int_0^1\frac{x(1-x)g^2(1-2x)dx}{\left[m^2-x(1-x)M^2\right]^3}\ .
\end{equation}

The elastic electromagnetic form factors can be also expressed via
$g(z)$  \cite{dshvk}:
\begin{eqnarray}\label{ff}
 F(Q) &=&
\frac{1}{8\pi^2N_{tot}}\int_0^1g(1-2x)dx\int_0^1g(1-2x')dx'
\nonumber\\
&\times&\int_0^1du \;u^2(1-u)^2 \frac{[xx'u(1-u)Q^2+(6\xi-5)m^2
+2M^2\xi(1-\xi)]}{[xx'u(1-u)Q^2+m^2-\xi(1-\xi)M^2]^4},
\end{eqnarray}
where $\xi=(1-x)u+(1-x')(1-u)$ and $Q^2=-q^2=-(p-p')^2>0$.
The constant $N_{tot}$ is found from the condition $F(0)=1$.
The transition form factor (between two different states) was
calculated according to ref. \cite{tff}. 

%%%%%%%%%%%%%%%%%%%%%%%%%%%%%%%%%%
\section{Numerical results}\label{num}
\begin{figure}[!ht]
\begin{center}
\epsfxsize=4cm
\epsfysize=2.5cm
\epsfbox{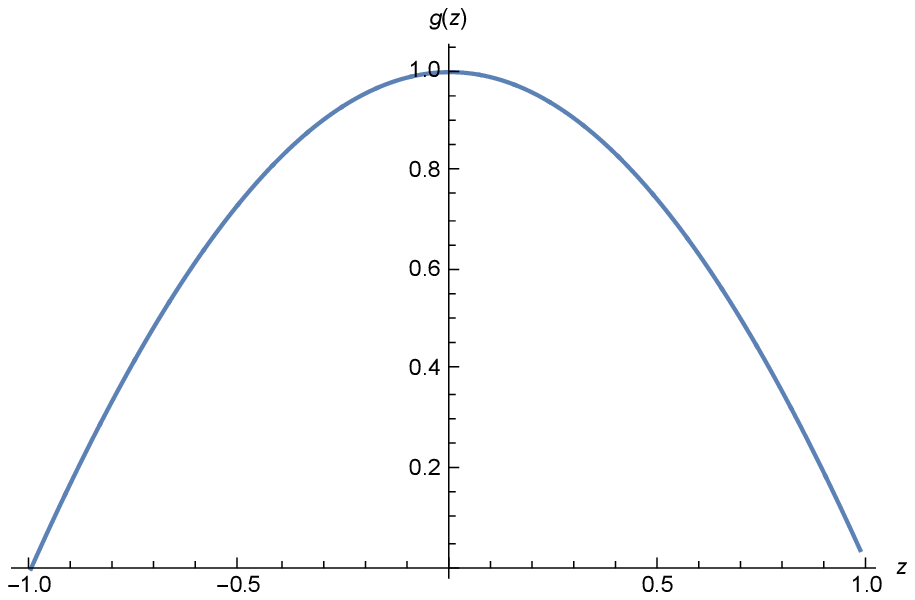}\hspace{0.5cm}
\epsfxsize=4cm
\epsfysize=2.5cm
\epsfbox{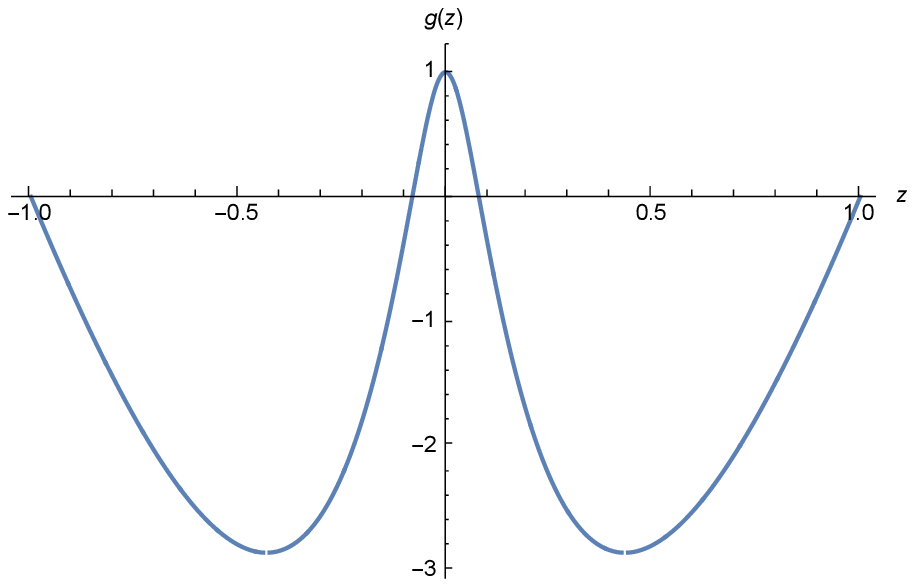}\hspace{0.5cm}
\epsfxsize=4cm
\epsfysize=2.5cm
\epsfbox{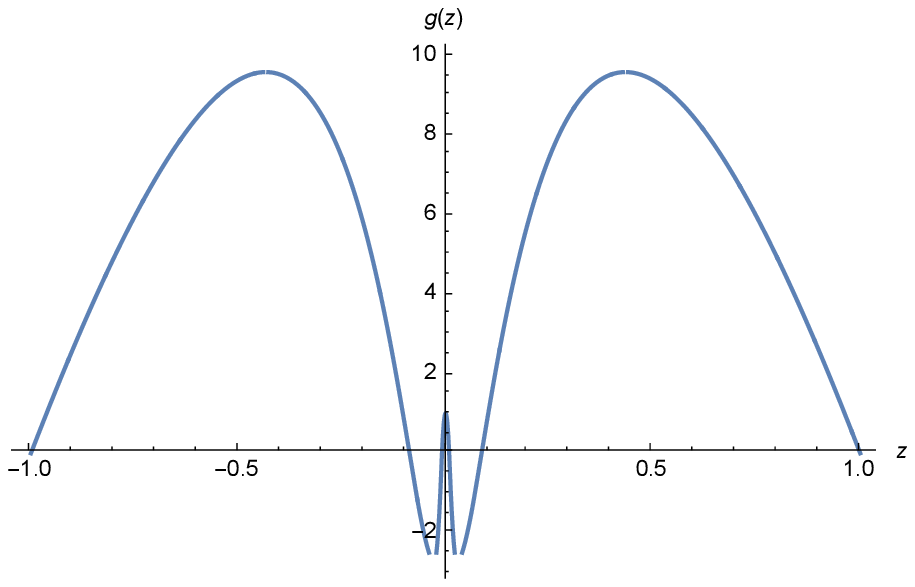}

\caption{The solutions $g(z)$ (normalized by $g(0)=1$) of Eq.
(\protect{\ref{g}}) for $\alpha=5$, $n=1$ and $k=0$ (at the left,
normal, zero nodes), $k=2$ (in the middle, abnormal, two nodes) and
$k=4$ (at the \mbox{right}, abnormal, four nodes).} 

\label{fig1}
\end{center}
\end{figure}

We solved numerically Eq. (\ref{g}) for $g(z)$, taking $m=1$, for a
few values of $\alpha$ in the interval $\alpha=0.02\div 5$. As an
example, in Fig. \ref{fig1} the solutions $g(z)$ for $\alpha=5$, $n=1$
and $k=0,2,4$ are shown.
\vspace{-0.5cm}

Using these solutions, we calculated by Eq. (\ref{norm2}) the
two-body contributions $N_2$ to the total norm  for a few low-lying
states with $n=1$, both normal ($k=0$) and symmetric abnormal
($k=2,4$). The results are given in Table \ref{tab1}.  For the
abnormal states they are shown in \textcolor{red}{red}.

\begin{table}[!ht]
\begin{center}
\begin{tabular}{clcll}
\hline\noalign{\smallskip} 
 No.&      $\alpha$ & $k=N_{nodes}$ &  $B$ & $N_2$\\
\noalign{\smallskip}
\hline
\noalign{\smallskip}
1&0.02    & 0   &   $10^{-4}$ &       0.992   \\
2&0.2   &0&   0.01          & 0.931    \\
3&2 &0  &   0.236    &  0.695 \\
\textcolor{red}{4}&\textcolor{red}{2}  &\textcolor{red}{2}&
\textcolor{red}{$1.220\cdot 10^{-5}$}
& \textcolor{red}{$7.7\cdot 10^{-3}$}    \\
5&5 &0  &   0.999         &0.651    \\
\textcolor{red}{6}&\textcolor{red}{5} &\textcolor{red}{2 }&
\textcolor{red}{$3.512 \cdot 10^{-3}$}
&  \textcolor{red}{$9.35\cdot 10^{-2}$}  \\
\textcolor{red}{7}&\textcolor{red}{5}  &\textcolor{red}{4}&
\textcolor{red}{$0.217\cdot 10^{-4}$}
&  \textcolor{red}{$8.55\cdot 10^{-3}$} \\
\noalign{\smallskip}\hline
\end{tabular}
\end{center}
\caption{Two-body contributions $N_2$ into the full normalization
integral (equal to 1) for some coupling constants $\alpha$ and corresponding
binding energies $B=2m-M$. The states Nos. 5 (normal), \textcolor{red}{6} and \textcolor{red}{7} (both abnormal)
correspond to the solutions shown in Fig. \ref{fig1}.
\label{tab1}}
\end{table}

For the normal states with $\alpha\ll 1$, the binding energy $B$
coincides with the Coulombic value $\frac{1}{4}\alpha^2$.
Besides, the values of $N_2$ vary in the "normal" interval  between
(for small $\alpha=10^{-4}$) $N_2=0.99$ (the limiting case
$N_2(\alpha\to 0)\to 1$) and (for large $\alpha=5$) $N_2=0.65$
(the limiting case $N_2(\alpha\to2\pi)\to 0.64$).

Whereas, for the abnormal states ($k=N_{nodes}=2$ and 4), even with
very small binding energies, the two-body contribution is of the order
of a few per cent only. This means that the abnormal states are mainly
many-body ones (two constituents with mass $m$ + a few or many
exchanged massless particles). This explains their disappearance
in the non-relativistic limit which corresponds to the bound state of two 
constituent particles only. In this limit, exchange by massless particles 
generates the Coulomb potential but does not generate their contribution 
to the state vector.

\begin{figure}[!ht]
\begin{center}
\epsfxsize=4cm
\epsfysize=2.5cm
\epsfbox{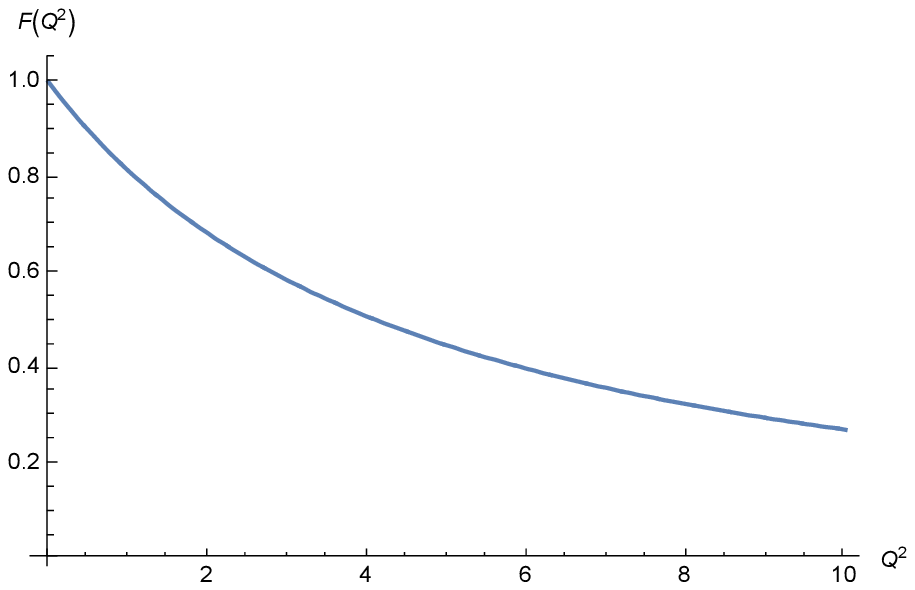}\hspace{0.5cm}
\epsfxsize=4cm
\epsfysize=2.5cm
\epsfbox{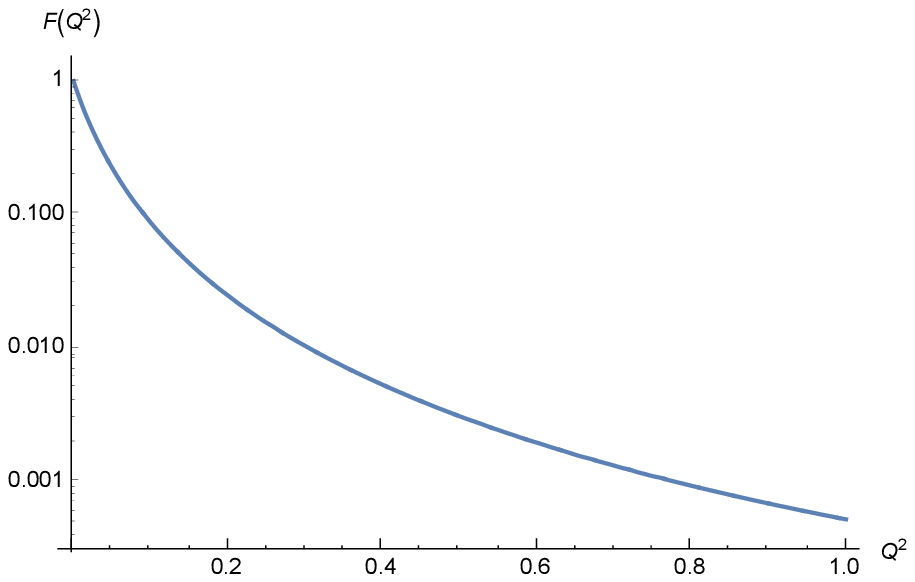}\hspace{0.5cm}
\epsfxsize=4cm
\epsfysize=2.5cm
\epsfbox{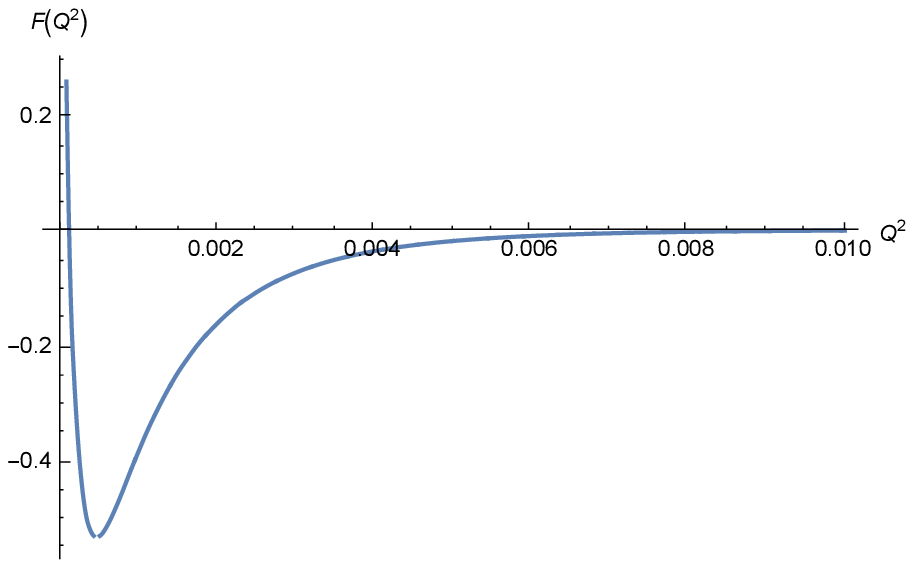}
\end{center}

\caption{Left, middle and right are the elastic form factors of the
states No. 5 (normal), \textcolor{red}{6} and \textcolor{red}{7} (both abnormal) 
from the Table \protect{\ref{tab1}} (calculated correspondingly with
left, middle and right $g(z)$ in Fig. \protect{\ref{fig1}}).} 
\label{fig2}

\end{figure}

Figure \ref{fig2} shows that the elastic em form factors of abnormal
states (middle and right) vs. $Q^2$ decrease, at least, $\sim 1000$
times faster than those of the normal states (left), as should be for
a many-body system.

\begin{figure}[!ht]
\begin{center}
\epsfxsize=4cm
\epsfysize=2.5cm
\epsfbox{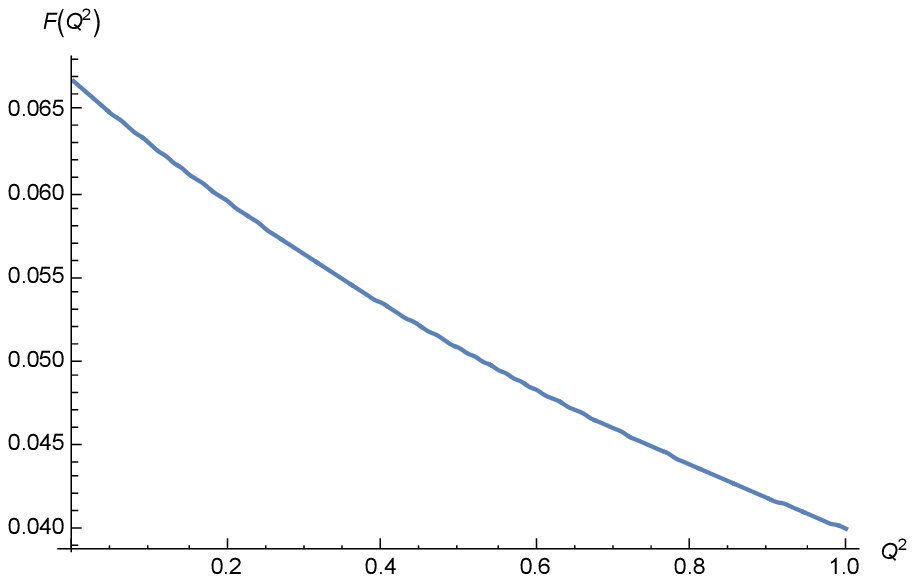}\hspace{0.5cm}
\epsfxsize=4cm
\epsfysize=2.5cm
\epsfbox{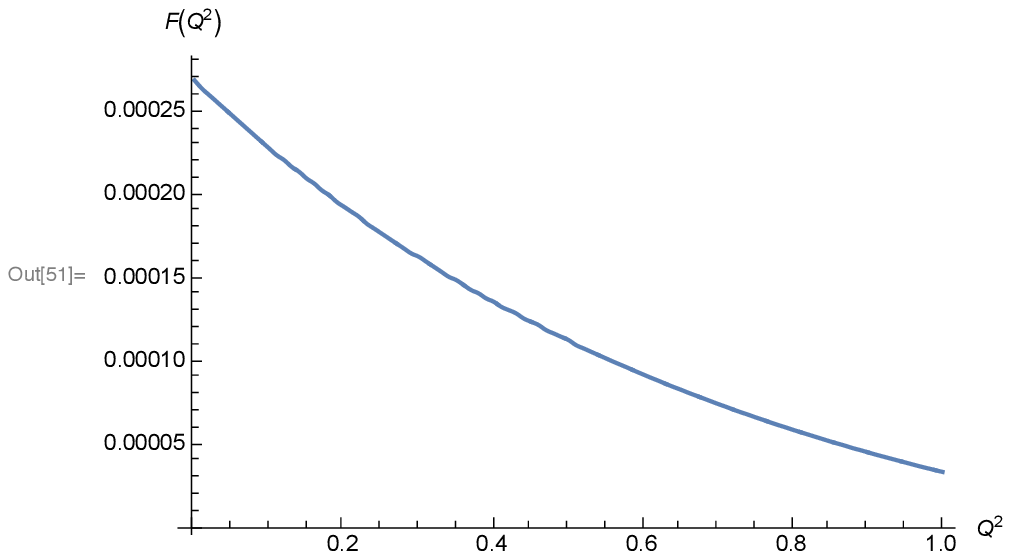}\hspace{0.5cm}
\epsfxsize=4cm
\epsfysize=2.5cm
\epsfbox{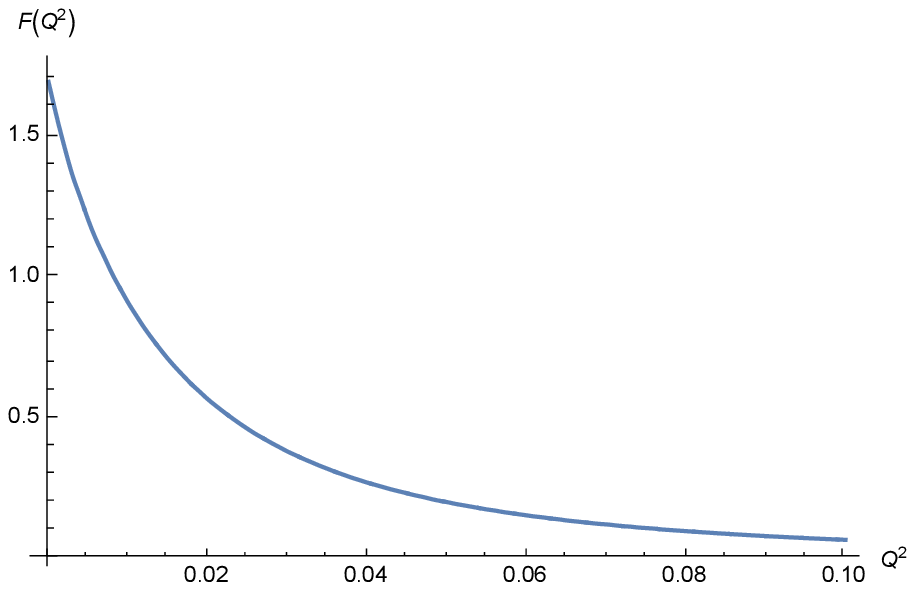}
\caption{Left is the transition form factor between the (normal) state
No. 5 from the Table \protect{\ref{tab1}} and the first abnormal one No. \textcolor{red}{6}; middle is the transition form
factor between the (normal) state No. 5 and the second abnormal one
No. \textcolor{red}{7}; right is the transition form factor between the two abnormal
states No. \textcolor{red}{6} and  \textcolor{red}{7}.
\label{fig3}}
\end{center}
\end{figure}

Figure \ref{fig3} shows  that the transition form factors between the
normal and abnormal states are a few orders of magnitude smaller than
those between the abnormal ones. In this sense, the normal and abnormal
states represent different "worlds", weakly "communicating" with each
other.

%%%%%%%%%%%%%%%%%%%%%%%%%%%
\section{Conclusion}\label{concl}

The normal and abnormal states of the Bethe-Salpeter equation drastically differ by their Fock-space content.
In the normal states the two-body (constituent) contribution dominates,
whereas the abnormal ones are dominated by the many-body contributions
of the exchanged particles.

These observations shed light on the nature and properties of the
abnormal states.

%%%%%%%%%%%%%%%%%%%%%%%%%%%%%%%%%%%%

\end{document}